\documentclass[fleqn,usenatbib]{mnras}

\usepackage{newtxtext,newtxmath}

\usepackage[T1]{fontenc}

\DeclareRobustCommand{\VAN}[3]{#2}
\let\VANthebibliography\thebibliography
\def\thebibliography{\DeclareRobustCommand{\VAN}[3]{##3}\VANthebibliography}

\usepackage{graphicx}	
\usepackage{amsmath}	
\usepackage{multirow}
\usepackage{hyperref}
\usepackage{bm}
\usepackage{latexsym}
\usepackage{epsfig}
\usepackage{psfrag}
\usepackage{ulem}
\usepackage{color}
\usepackage[dvipsnames]{xcolor}
\usepackage{subfigure}
\usepackage[utf8]{inputenc}

\title[Neutrino mass constraints in 4pDE]{Neutrino mass constraints in the context of 4-parameter dark energy equation of state and DESI DR2 observations}

\author[Gowri S Nair et al.]{
Gowri S Nair,$^{1,2}$\thanks{gowrisnair22@iisertvm.ac.in}
Amlan Chakraborty,$^{2,3}$ \thanks{amlan.chakraborty@iiap.res.in}
Luca Amendola$^{3,4}$ \thanks{l.amendola@thphys.uni-heidelberg.de}
and Subinoy Das$^{2,3}$ \thanks{subinoy@iiap.res.in}
\\
$^{1}$School of Physics, Indian Institute of Science Education and Research Thiruvananthapuram, Maruthamala PO, Vithura, Thiruvananthapuram 695551, Kerala, India\\
$^{2}$Indian Institute of Astrophysics, Bengaluru, Karnataka 560034, India\\
$^{3}$Institut f\"ur Theoretische Physik, Universit\"at Heidelberg,
Philosophenweg 16, 69120 Heidelberg, Germany\\
$^{4}$New York University Abu Dhabi, PO Box 129188, Abu Dhabi, United Arab Emirates and Center for Astrophysics and Space Science (CASS), New York University \\ Abu Dhabi
}

\date{Accepted XXX. Received YYY; in original form ZZZ}

\pubyear{\the\year{}}

\begin{document}
\label{firstpage}
\pagerange{\pageref{firstpage}--\pageref{lastpage}}
\maketitle

\begin{abstract}
Cosmological constraints on the total neutrino mass, $\sum m_\nu$, are strongly shaped by assumptions about the dark-energy equation of state due to the well-known degeneracy between massive neutrinos and late-time cosmic acceleration. In this work, we move beyond the two-parameter Chevallier–Polarski–Linder (CPL) form adopted in recent DESI analyses and re-examine neutrino mass constraints using a flexible four-parameter dark energy equation of state (4pDE). We implement the 4pDE model in a modified version of CLASS and perform a full MCMC analysis using Planck, DESI DR2 BAO, and Pantheon+ data. Relative to our previous 4pDE study based on pre-DESI BAO datasets, the inclusion of DESI DR2 substantially tightens the constraints on the transition parameters while still yielding a relaxed neutrino-mass bound compared to $\Lambda$CDM, $\sum m_\nu < 0.101$ eV ($95\%$ C.L.). This upper limit is more stringent than the DESI DR2 constraint obtained within the $w_0w_a$CDM framework. From the best-fit parameters, we reconstruct the evolution of the 4pDE equation of state along with both $68\%$ and $95\%$C.L. We do not find a statistically significant phantom-crossing at $z \sim 0.5$, consistent with the conclusion from the DESI collaboration; at higher redshifts, the reconstructed $w(z)$ follows the CPL evolution and deviates only at low redshift. Additionally we also find reduction in $\Delta \chi^2_{\rm min}=-7.3$ compared to $\Lambda$CDM model.
\end{abstract}




\section{Introduction}\label{sec:intro}

The absolute scale of neutrino masses remains one of the key open questions at the interface of particle physics and cosmology. 
Oscillation experiments have firmly established that neutrinos possess mass splittings, but the total mass scale, $\sum m_\nu$, is only weakly constrained. 
Cosmological observables—especially the cosmic microwave background (CMB), baryon acoustic oscillations (BAO), and large-scale clustering—offer the most sensitive indirect probe of $\sum m_\nu$ \citep{LesgourguesPastor2012,HuEisenstein1998,Lattanzi:2017ubx, Wang:2025ker, Bertolez-Martinez:2024wez, Wang:2025zuo, Sabogal:2025qhz}.  

A well-known feature of cosmological parameter inference is the degeneracy between the total neutrino mass $\sum m_\nu$ and the dark-energy equation of state parameters. 
Massive neutrinos suppress the growth of structure and shift the expansion rate, both of which can be partially compensated by changes in the dark-energy sector. 
For example, increasing $\sum m_\nu$ lowers the late-time amplitude of matter clustering, but a less negative $w$ (i.e., $w > -1$) can produce a similar integrated effect on the growth factor and distance measures. 
As a result, cosmological observables such as BAO and CMB lensing cannot independently distinguish between the impact of neutrino mass and that of evolving dark energy without external priors \citep{Hannestad2005,Vagnozzi2018, Elbers:2024sha, Sharma:2025iux,GarciaEscudero:2025lef}.  

This degeneracy implies that the inferred upper limit on $\sum m_\nu$ depends on the assumed functional form of $w(z)$. 
When dark energy is fixed to a cosmological constant, the expansion history is rigidly determined, leading to tight bounds on $\sum m_\nu$.  
Allowing $w(z)$ to vary—as in the two-parameter CPL model or higher-dimensional extensions which introduce additional freedom in late-time acceleration can absorb part of the neutrino-induced suppression of structure growth \citep{Sharma:2022ifr,Du:2024pai,Elbers:2024sha}.  
The recent result from  \textit{Dark Energy Spectroscopic Instrument (DESI)} has ushered in a new era of high-precision BAO measurements \citep{DESI2024BAO,Adame2024DESI,Elbers:2025vlz}.  
DESI Data Release~1, together with subsequent DESI data releases, has already surpassed the precision of BOSS/eBOSS and constrained $\sum m_\nu \lesssim 0.07$~eV (95\% C.L.) within $\Lambda$CDM.
Such tight limits motivate careful examination of model assumptions, since cosmological bounds on $\sum m_\nu$ are known to be degenerate with dark-energy dynamics \citep{Hannestad2005,Ichikawa2005,dePutterLinder2008,Vagnozzi2018,RoyChoudhury:2025dhe,Jiang:2024viw}.

Dark energy is commonly described by the two-parameter Chevallier–Polarski–Linder (CPL) form $w(a)=w_0+w_a(1-a)$ \citep{Chevallier2001,Linder2003}, 
which captures first-order deviations from a cosmological constant and fits current data well \citep{DESI:2025zgx, DESI2024BAO,DESI:2025fii,Zhou:2025nkb, Du:2025xes,Reeves:2025xau}.  
However, a two-parameter evolution to fit the data might miss a richer evolution of $w(z)$ as well as non-trivial theoretical models arising from 
quintessence potential including phantom crossing and transitions in $w(z)$ \citep{Caldwell2002,DeFeliceTsujikawa2010,PhysRevD.111.083547,Nesseris:2025lke,Lee:2025ysg,Artola:2025zzb,Feng:2025mlo, Du:2025iow,Barua:2025adv}.  
That is why multi-parameter or non-parametric extensions are essential, \cite{Giare:2024gpk}, to test the robustness of neutrino-mass bounds.

 Previous study, \citep{Sharma:2022ifr}, introduced a four-parameter dark-energy equation of state (4pDE) for a robust evolution of dark energy equation of state.  That work demonstrated that extended EoS freedom can shift $\sum m_\nu$ constraints and alter degeneracy directions. That analysis was performed before the DESI data were released. 
Several subsequent investigations explored similar ideas, showing that generalized $w(z)$ forms can relax neutrino-mass bounds while maintaining consistency with CMB and SN data \citep{Wang2022PRD,Yao2023JHEP,Ren2023PRD,Zhao2023EPJC}.
DESI’s exquisite BAO precision now allows these degeneracies to be revisited quantitatively.
Recent work by V. Miranda et.al \cite{Miranda2023JCAP,Reboucas:2024smm} performed a comprehensive joint analysis of DESI Year 1 BAO, Planck, and SN data, 
highlighting how dynamical dark-energy models can modify cosmological neutrino-mass limits.  
Similar conclusions have been drawn by \cite{Du:2024pai,Elbers:2024sha,Rodrigues:2025zvq, Chebat:2025kes}, 
who emphasized that allowing a flexible late-time EoS can mimic or mask small neutrino-mass signals.

In this work, we re-examine the neutrino-mass constraints using our 4pDE model in light of the latest DESI BAO results.  
We perform a thorough Markov-Chain Monte-Carlo (MCMC) analysis with the \texttt{COBAYA} framework \citep{Torrado:2020dgo}, 
jointly fitting $\sum m_\nu$, standard cosmological parameters, and the four dark-energy coefficients using Planck, DESI DR2 BAO, and Pantheon+ SN-Ia data. Compared to our previous analysis with pre-DESI BAO datasets, the inclusion of DESI data tightens the allowed parameter space and breaks degeneracies between $\sum m_\nu$ and dark-energy evolution. This improvement arises because the improved precision of DESI over BAO distance measurements provides strong constraints on the expansion history at intermediate redshifts, precisely where variations in $w(z)$ can mimic the effects of massive neutrinos. This leads to a reduction in the freedom of the 4pDE evolution to compensate for neutrino-induced suppression of structure growth, thereby breaking the $\sum m_\nu$–$w(z)$ degeneracy and shrinking the allowed parameter volume.

From our MCMC results, we reconstruct the redshift-dependent equation of state $w(z)$ from the obtained best-fit values of the parameters and find that the data prefer a clear phantom crossing around $z\sim0.5$ when the 4pDE model is adopted.  
Because the reconstructed $w(z)$ differs markedly from $\Lambda$CDM at low redshift, the inferred upper bound on $\sum m_\nu$ becomes relaxed relative to the baseline $\Lambda$CDM result—consistent with the trend reported by \citep{Vagnozzi2018,Du:2024pai}.  
Our findings highlight the crucial interplay between dark-energy modeling and cosmological neutrino-mass inference.

In this work, using a joint dataset comprising Planck, DESI DR2 BAO, and Pantheon+, we find that the upper bound on the total neutrino mass is relaxed relative to the standard $\Lambda$CDM constraint, yielding $\sum m_\nu < 0.101~\mathrm{eV}\;(95\% \text{C.L.})$. However, this limit remains tighter than both our previous result \cite{Sharma:2022ifr} and the recent DESI collaboration analysis employing the CPL parameterization. Because of the well-known degeneracy between $\sum m_\nu$ and the time evolution of the dark-energy equation of state, a more flexible model such as our four-parameter 4pDE framework is better equipped to capture variations in $w(z)$ than the two-parameter CPL form. Although not all four coefficients of the 4pDE model are individually well constrained, the high-precision DESI~DR2 BAO measurements—significantly more precise than pre-DESI data—strongly limit the combinations of $w(z)$ that mimic neutrino-induced effects. This effectively breaks the $\sum m_\nu$–$w(z)$ degeneracy and leads to a tighter neutrino-mass bound in the 4pDE model than in the CPL case, despite the additional degrees of freedom. In addition, our model provides a more stringent constraint on one additional cosmological parameter compared to our earlier work.

This paper is structured as follows. Section~\ref{sec:model} reviews the 4pDE model and its motivation. Section~\ref{sec:data} describes the data sets and our MCMC methodology. Section~\ref{sec:results} presents constraints on the dark-energy parameters, reconstructed $w(z)$, and derived bounds on $\sum m_\nu$. Section~\ref{sec:conclusion} discusses the implications of our results for upcoming DESI data releases and next-generation surveys.

\section{Model} \label{sec:model}

In many particle-physics–motivated dark energy (DE) models, the commonly used Chevallier–Polarski–Linder (CPL) parametrization fails to accurately capture the dynamical evolution of the dark energy equation-of-state (EoS) parameter. To achieve a more faithful description, it becomes necessary to extend the parameter space beyond the conventional two-parameter form. In this work, we adopt a model-independent four-parameter EoS for dark energy, following the formulation introduced in \cite{Sharma:2022ifr}. The EoS is expressed as

\begin{equation}
\begin{aligned}
w_{\rm de}(a) =\; & w_0 
+ (w_m - w_0)\,
\frac{1 - \exp\!\left[-\dfrac{a - 1}{\Delta_{\rm de}}\right]}
     {1 - \exp\!\left[-\dfrac{1}{\Delta_{\rm de}}\right]}\,
\frac{1 + \exp\!\left[ \dfrac{a_{\rm t}}{\Delta_{\rm de}} \right]}
     {1 + \exp\!\left[-\dfrac{a - a_{\rm t}}{\Delta_{\rm de}} \right]} .
\end{aligned}
\label{eq:de_eqn_state}
\end{equation}

where the four parameters $w_0$, $w_m$, $a_{\rm t}$, and $\Delta_{\rm de}$ control the evolution of the dark energy equation of state. Here, $w_0$ denotes the value of the equation of state parameter at the present epoch, i.e., $a=1$, while $w_m$ corresponds to its asymptotic value in the early Universe $(a\ll1)$. $a_{\rm t}$ is the scale factor at which the transition between the two asymptotic regimes occurs, and $\Delta_{\rm de}$ characterizes the smoothness or duration of this transition.  

This formulation allows a smooth but flexible evolution of the dark-energy equation of state that naturally encompasses a wide range of behaviors—from early-time freezing to late-time thawing models—without committing to a specific underlying scalar-field potential. The model also recovers familiar forms as special limits: a constant equation of state ($w_0=w_m$) or the standard two-parameter CPL form when the transition is sufficiently broad and monotonic.

Correspondingly, the Friedmann equation becomes,

\begin{equation}
\begin{aligned}
3H^{2} M_{\rm Pl}^{2} =\;& 
\rho_{\gamma}^{0} a^{-4}
+ \rho_{\rm cdm}^{0} a^{-3}
+ \rho_{\rm b}^{0} a^{-3} \\[0.5ex]
& +\, \rho_{\rm de}^{0}
\exp\!\left[
3 \int_{1}^{a} \!\bigl(1 + w_{\rm de}(a')\bigr)\, d\ln a'
\right] 
\end{aligned}
\label{eq:friedmann}
\end{equation}

where $\rho_\gamma^0, \rho_{\rm cdm}^0, \rho_{\rm b}^0, \rho_{\rm de}^0$ are the present-day energy densities of photon, cold dark matter, baryon, and dark energy, respectively, and $M_{\rm Pl}$ is the Planck mass. 

This model can also potentially leave distinct and complementary signatures across different cosmological observables. Although dark energy becomes the dominant component at late times, its evolving equation of state influences the cosmic microwave background (CMB) primarily through geometric and late-time effects. Changes in the background expansion modify the angular-diameter distance to the surface of last scattering, shifting the locations of the acoustic peaks in the temperature and polarization spectra. Also, the time variation of the gravitational potential can affect the late Integrated Sachs–Wolfe (ISW) signal at large angular scales. The baryon acoustic oscillation (BAO) measurements, which act as a standard ruler, are directly sensitive to the expansion history (Equation \ref{eq:friedmann}), which is directly influenced by any evolution in $w_{\rm de}(a)$, as shown in Eq. \ref{eq:de_eqn_state}. Variations in the dark-energy parameters alter the comoving sound horizon at the baryon-drag epoch and thereby shift the inferred distances from galaxy clustering. Finally, Type Ia supernovae probe the luminosity-distance $d_L(z)$, providing purely geometric constraints on the late-time expansion history. Because $d_L(z)\propto \int dz'/H(z')$ (in flat space), even mild departures of $w_{\rm de}(a)$ from $-1$ can produce measurable shifts in the distance modulus at intermediate redshifts. Taken together, CMB, BAO, and SNe Ia data can thus probe the model by providing robust joint constraints on the parameters $\{w_0, w_m, a_{\rm t}, \Delta_{\rm de}\}$ and their degeneracies with the neutrino mass.

\section{Method}\label{sec:data}

We modified the Boltzmann solver \texttt{CLASS} to incorporate and evolve the full set of background and perturbation equations corresponding to the 4pDE model, following the formulation outlined in~\cite{Sharma:2022ifr}. This implementation enables the computation of the cosmic microwave background (CMB) temperature and polarization anisotropy spectra, as well as the linear matter power spectrum, across the model’s parameter space.

A comprehensive Markov Chain Monte Carlo (MCMC) analysis is performed using the following cosmological datasets to constrain the model parameters:

\begin{itemize}
    \item[$\blacktriangleright$] \textbf{Planck 2018}: Temperature (TT), polarization (EE), and cross-correlation (TE) power spectrum measurements from Planck, employing the low-$\ell$ \texttt{simall, Commander} likelihood for $\ell < 30$ and the high-$\ell$ \texttt{CamSpec} likelihood for $\ell \ge 30$~\cite{refId0}.
    \item[$\blacktriangleright$] \textbf{DESI DR2 BAO}: Baryon Acoustic Oscillation (BAO) measurements from the second data release of DESI~\cite{DESI:2025zgx}, which include contributions from multiple tracers—galaxies, quasars, and the Lyman-$\alpha$forest. These data provide both isotropic and anisotropic distance constraints over the redshift interval $0.295 \leq z \leq 2.330$, subdivided into nine redshift bins as described in Table~IV of~\cite{DESI:2025zgx}.
    \item[$\blacktriangleright$] \textbf{Pantheon$+$}: $1550$ Unique Type Ia supernovae measurements spanning over a redshift range of $z = 0.001$ to $2.26$.\cite{Brout_2022}
    \item[$\blacktriangleright$] \textbf{Planck PR4 lensing}: High-$\ell$ temperature (TT), temperature–polarization (TE), and polarization (EE) power spectra derived from the NPIPE maps, analyzed using the \texttt{CamSpec} likelihood, together with the gravitational lensing potential power spectrum ($C_L^{\phi\phi}$) reconstructed from the PR4 dataset~\cite{Carron:2022eyg}.

\end{itemize}
Our baseline cosmological model includes the six standard $\Lambda$CDM parameters: 
$\{\omega_{\rm b},\:\omega_{\rm dm},\:\theta_s,\:\ln(10^{10}A_s),\:n_s,\:\text{and}\:\tau_{\rm reio}\}\,,$ in addition to the four  Dark Energy  parameters, $\{w_0$, $w_m$, $a_{\rm t},\; \Delta_{\rm de}\}$ and total neutrino mass $\sum m_\nu$. 

The Markov Chain Monte Carlo (MCMC) analysis is carried out using the publicly available framework \texttt{COBAYA}~\cite{Torrado:2020dgo}, which implements the Metropolis–Hastings algorithm and interfaces directly with our modified version of \texttt{CLASS}. Sampling efficiency is enhanced through Cholesky decomposition of the covariance matrix, allowing for an optimized exploration of fast and slow directions in the parameter space~\cite{Lewis_2000}. Convergence of the MCMC chains is assessed using the Gelman–Rubin diagnostic, requiring $R-1 < 0.01$ for all sampled parameters~\cite{gelman1992inference}.

To assess the statistical preference between models, we compute the Akaike Information Criterion (AIC). The difference in AIC values between the 4pDE model and the reference $\Lambda$CDM model is defined as \cite{1100705}
\begin{equation}\label{eq:AIC}
\Delta\mathrm{AIC} = \chi^2_{\mathrm{min,4pDE}} - \chi^2_{\mathrm{min,\Lambda CDM}} + 2(N_{\mathrm{4pDE}} - N_{\mathrm{\Lambda CDM}})\, .
\end{equation}
A negative $\Delta\mathrm{AIC}$ implies that the extended dark-energy model provides a better statistical fit to the data than $\Lambda$CDM for the same dataset combination, whereas a positive value indicates the opposite. Following the convention in \cite{mcgb-ntwr}, we interpret the strength of evidence as follows: weak ($-2 \leq \Delta\mathrm{AIC} < 0$), positive ($-6 \leq \Delta\mathrm{AIC} < -2$), strong ($-10 \leq \Delta\mathrm{AIC} < -6$), and very strong ($\Delta\mathrm{AIC} < -10$) in favor of the model; conversely, weak ($0 < \Delta\mathrm{AIC} \leq 2$), positive ($2 < \Delta\mathrm{AIC} \leq 6$), strong ($6 < \Delta\mathrm{AIC} \leq 10$), and very strong ($\Delta\mathrm{AIC} > 10$) against it.

\section{Results}\label{sec:results}

\begin{figure*}
    \centering
    \includegraphics[width=1.0\linewidth]{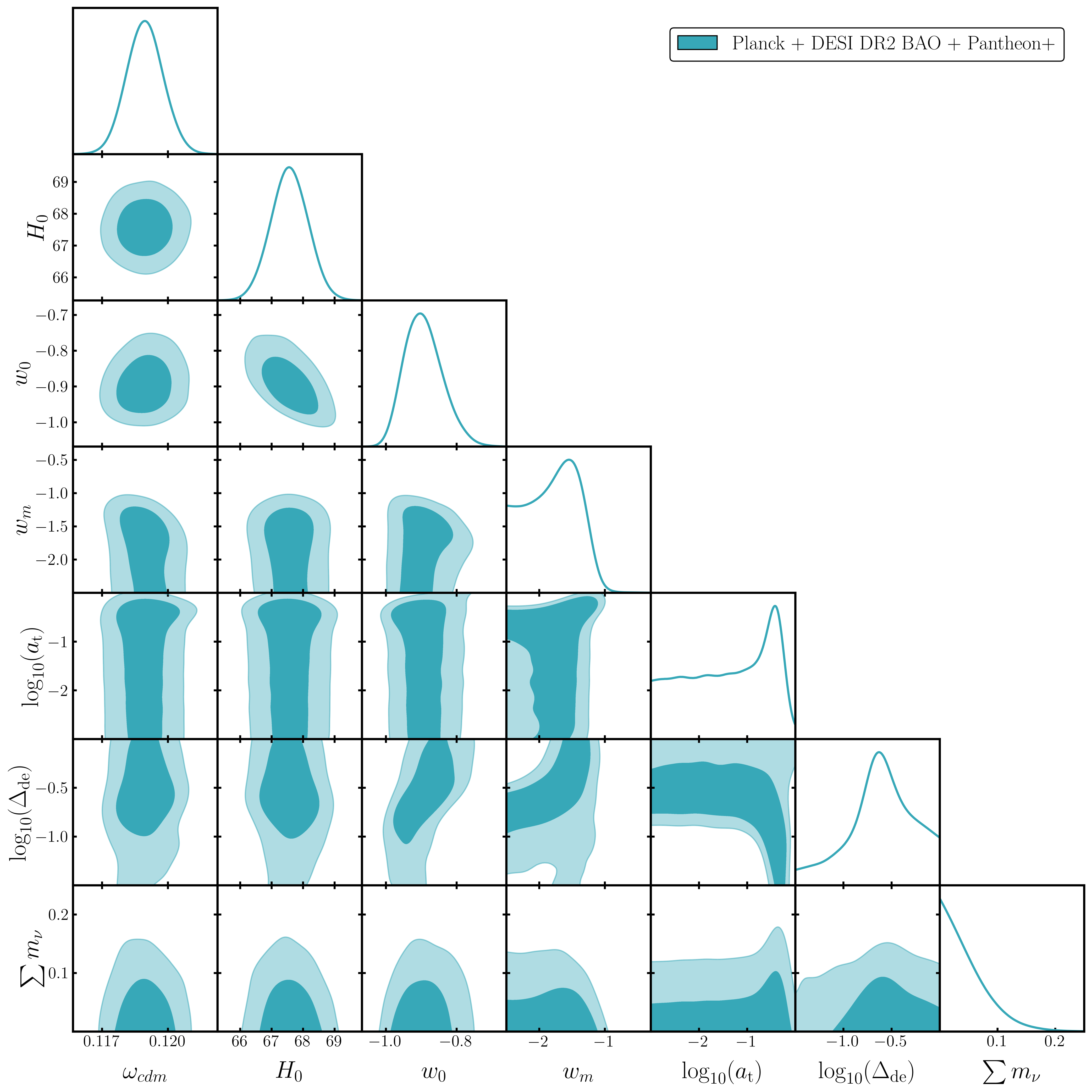}
    \caption{
    Two-dimensional marginalized posterior distributions of the 4pDE model parameters ($\omega_{\rm cdm}$, $H_0$, $w_0$, $w_m$, $\log_{10} (a_{\rm t})$, $\log_{10} (\Delta_{\rm de})$, and $\sum m_\nu$) obtained from our MCMC analysis for the dataset combinations of Planck, DESI DR2 BAO and Pantheon+. The contours correspond to the 68\% and 95\% confidence regions.}
    \label{fig:contour}
\end{figure*}

\begin{figure*}
    \centering
    \includegraphics[width=0.9\linewidth]{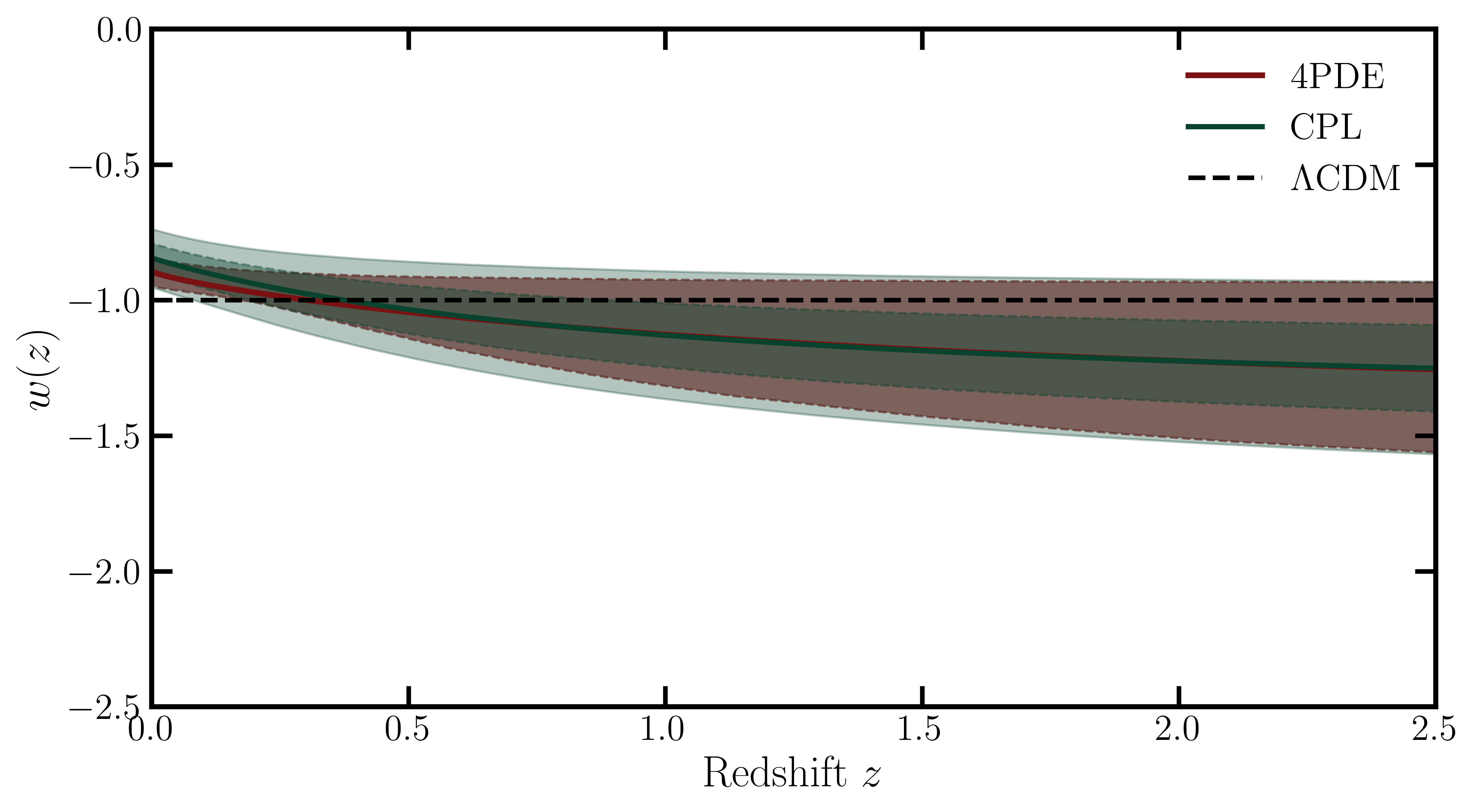}
    \caption{Dark energy equation of state plotted for the 4pDE model using the best-fit values obtained from MCMC analysis using Planck+DESI DR2+Pantheon+. The shaded regions indicate the obtained $1\sigma$ and $2\sigma$ confidence regions. The result is then compared with CPL parameterization obtained from DESI analysis \protect\citep{DESI:2025zgx}}
    \label{fig:reconstruction}
\end{figure*}

We perform two independent MCMC analyses using the combined dataset of Planck, DESI DR2 BAO, and Pantheon+ for both the standard $\Lambda$CDM model and the extended four-parameter dark-energy model (4pDE). For both cases, we keep the sum of the neutrino mass always positive as a prior. The resulting constraints for the 4pDE case are summarized in Table~\ref{Tab:Tab1}, while Figure~\ref{fig:contour} displays the corresponding marginalized one- and two-dimensional posterior distributions. The reconstructed dark energy equation of state $w(z)$, inferred from the best-fit parameters together with the $1\sigma$ and $2\sigma$ confidence regions, is shown in Figure~\ref{fig:reconstruction}. 

Figure~\ref{fig:contour} demonstrates that the present-day equation of state parameter $w_0$ is tightly constrained by the combined dataset and is consistent with the DESI DR2 analysis obtained within the $w_0w_a$CDM parametrization~\cite{DESI:2025zgx}. The remaining dark-energy parameters—$w_m$, $\log_{10}(a_{\rm t})$, and $\log_{10}(\Delta_{\rm de})$—remain only weakly constrained, though the constraining power is notably improved relative to earlier 4pDE studies~\cite{Sharma:2022ifr}, particularly for $\log_{10}(\Delta_{\rm de})$ and the total neutrino mass $\sum m_\nu$. 

\begin{table*}
\centering
\Large
\setlength{\tabcolsep}{25pt}        
\renewcommand{\arraystretch}{1.3}  

\begin{tabular}{lcc}
\hline
\rule{0pt}{1.2\normalbaselineskip}%
 & \multicolumn{2}{c}{Planck + DESI DR2 BAO + Pantheon+} \\
\cline{2-3}
\rule{0pt}{1.2\normalbaselineskip}%
Parameters & $\Lambda$CDM & 4pDE \\
\hline


$100\,\omega_{\rm b}$ &
$2.25\,(2.26)\pm 0.0131$ &
$2.22\,(2.23)^{+0.0130}_{-0.0130}$ \\[1ex]

$\omega_{\rm cdm}$ &
$0.118\,(0.118)^{+0.000665}_{-0.000688}$ &
$0.1189\,(0.119)^{+0.00084}_{-0.00084}$ \\[1ex]

$100\,\theta_s$ &
$1.04\,(1.04)\pm 0.000278$ & $1.04\,(1.04)\pm 0.000281$
\\

$\ln(10^{10}A_s)$ &
$3.05\,(3.05)^{+0.0160}_{-0.0177}$ &
$3.036\,(3.036)^{+0.0139}_{-0.0140}$ \\[1ex]

$n_s$ &
$0.971\,(0.972)\pm 0.00342$ &
$0.9645\,(0.964)\pm 0.0037$ \\[1ex]

$\tau_{\rm reio}$ &
$0.0575\,(0.0580)^{+0.00764}_{-0.00845}$ &
$0.0532\,(0.0526)\pm 0.0073$ \\[1ex]

$\sum m_\nu$ [eV] &
$<0.0654$ &
$<0.101$ \\[1ex]

$w_0$ &
--- &
$-0.893\,(-0.917)^{+0.047}_{-0.060}$ \\[1ex]

$w_m$ &
--- &
$-1.79\,(-1.25)^{+0.48}_{-0.35}$ \\[1ex]

$\log_{10} (a_{\rm t})$ &
--- &
$<-1.8864$ \\[1ex]

$\log_{10} (\Delta_{\rm de})$ &
--- &
$-0.61\,(-1.32)^{+0.39}_{-0.26}$ \\[1ex]
\hline


$H_0$ &
$68.4\,(67.4)\pm 0.302$ &
$67.57\,(67.5)\pm 0.59$ \\[1ex]

$\sigma_8$ &
$0.805\,(0.832)^{+0.00687}_{-0.00751}$ &
$0.809\,(0.819)^{+0.0092}_{-0.0106}$ \\[1ex]
\hline

$\chi^2_{\min}$ &
$12389.3$ &
$12382$ \\[1ex]

$\Delta\chi^2_{\min}$ &
$0$ &
$-7.3$ \\
\hline
\end{tabular}

\vspace{1ex}
\caption{Mean (best-fit) $\pm1\sigma$ constraints for the $\Lambda$CDM and 4pDE models using
Planck + DESI DR2 BAO + Pantheon+. The $1\sigma$ errors are $68\%$ credible intervals.
We also quote $\chi^2_{\min}$ and $\Delta\chi^2_{\min}$. Upper bounds of the parameters are obtained at $95\%$ C.L.}
\label{Tab:Tab1}
\end{table*}

Figure~\ref{fig:reconstruction} illustrates the reconstructed dark energy equation of state derived from the best-fit parameters obtained through the MCMC analysis of the combined dataset. The shaded regions represent the $1\sigma$ and $2\sigma$ confidence intervals. The evolution of $w(z)$ closely mirrors the behavior reported by the DESI collaboration for the $w_0w_a$CDM model~\cite{DESI:2025fii}, exhibiting a characteristic phantom crossing around redshift $z \sim 0.5$ and transitioning to values above $-1$ at the present epoch.  

\begin{figure*}
    \centering
    \includegraphics[width=0.5\linewidth]{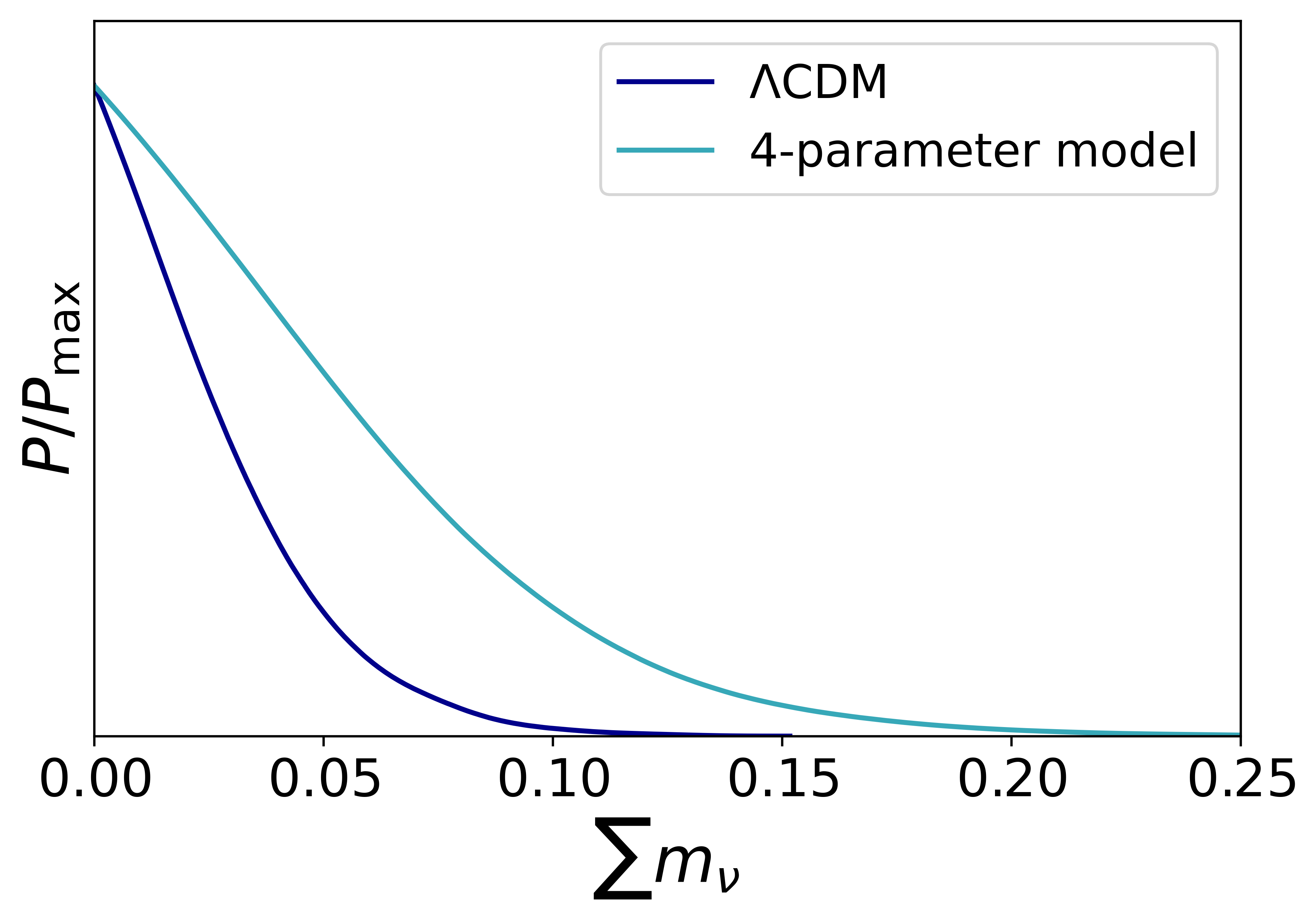}~
    \includegraphics[width=0.5\linewidth]{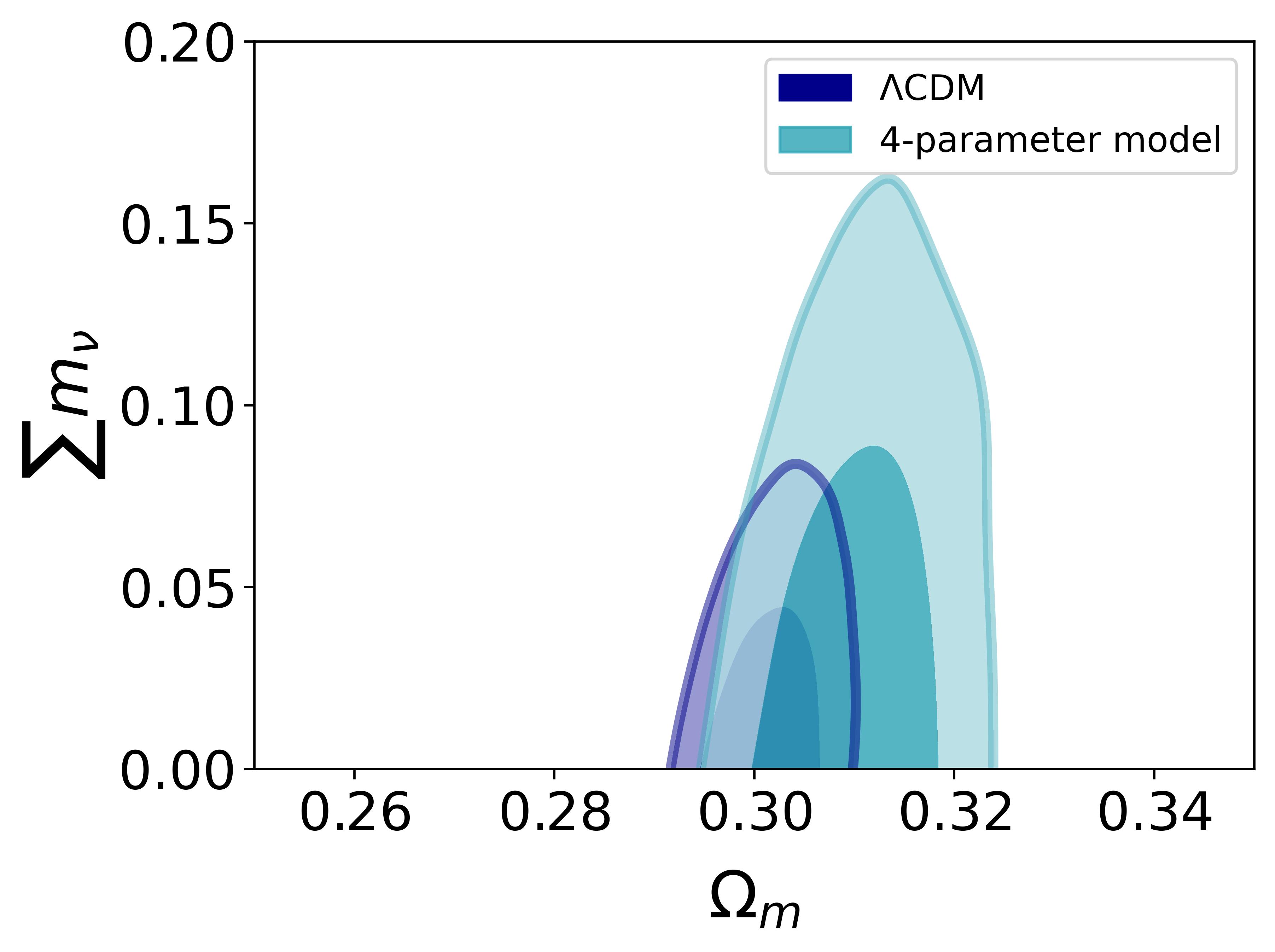}
    \caption{$1$D posterior distributions of the neutrino mass for the $\Lambda$CDM and 4pDE models (left), and the corresponding $2$D marginalized contours between the neutrino mass and $\Omega_m$ (right). All results are obtained from MCMC analyses using the combined Planck, DESI DR2 BAO, and Pantheon+ datasets.}
    \label{fig:m_nu_1d_and_cont}
\end{figure*}

The impact of the 4pDE model on the inferred neutrino mass is shown in Figure~\ref{fig:m_nu_1d_and_cont}, which presents the one-dimensional posterior distribution of $\sum m_\nu$ (left panel) and the two-dimensional contour in the $\sum m_\nu$–$\Omega_m$ plane (right panel). For the 4pDE model, we obtain a relaxed upper bound on the total neutrino mass of

\begin{equation*}
    \sum m_\nu < 0.101\; \text{eV}\; [95\% \;\text{C.L}]
\end{equation*}
compared to the tighter $\Lambda$CDM bound of

\begin{equation*}
    \sum m_\nu < 0.0654\; \text{eV}\; [95\% \;\text{C.L}]
\end{equation*}

Importantly, the 4pDE posterior shows no peak within the physically allowed (positive) mass range, implying that the maximum of the posterior lies in the prior-forbidden negative-mass region. This behavior is a well-known effect that arises when the posterior is prior-bounded at $\sum m_\nu=0$. Our result is consistent with constraining $\Lambda$CDM with only the Planck datasets \cite{refId0} as well as for the $w_0 w_a$CDM model used by the DESI collaboration \cite{DESI:2025zgx} for the same dataset combination as ours, i.e. Planck + DESI DR2 BAO + Pantheon+.   


When compared to earlier work on the same 4pDE model~\cite{Sharma:2022ifr}, which used pre-DESI era BAO datasets like SDSS DR7, BOSS DR12, eBOSS and Ly-$\alpha$ auto-correlation etc. and reported a $68\%$ bound of 
\begin{equation*}
    \sum m_\nu = 0.113_{-0.113}^{+0.0276}\; \text{eV}\,,
\end{equation*}
we see that the addition of DESI DR2 BAO leads to a substantially tighter upper bound on the total neutrino mass.
bound from our current analysis from 4pDE model, i.e. $\sum m_\nu < 0.101\; \text{eV}\; [95\% \;\text{C.L}]$, is more stringent than the constraint obtained by DESI collaboration within the $w_0w_a$CDM framework \cite{DESI:2025zgx} 
\begin{equation*}
    \sum m_\nu < 0.117\; \text{eV}\; [95\% \;\text{C.L}]
\end{equation*}
reflecting the increased constraining power of the joint dataset employed here. 
Note that the improvement with respect to pre-DESI data is even stronger, since earlier 4pDE constraints were given at $68\%$~C.L., whereas the DESI result and our analysis employ the more conservative $95\%$~C.L. Our 95\% limit is tighter than the previous 68\% bound, underscoring the significant gain in precision enabled by DESI’s high-quality BAO measurements and the extended redshift coverage of the joint dataset.

There is a well-established degeneracy between massive neutrinos and evolving dark energy, as both produce qualitatively similar effects on the matter power spectrum, contributing to the suppression and smoothing of small-scale density perturbations. In the 4pDE model, the additional dynamical freedom in $w(z)$ partially compensates the neutrino-induced power suppression, thereby reducing the ability of the data to isolate neutrino-mass effects and leading to a weaker upper limit on $\sum m_\nu$.

To statistically evaluate the model preference, we compute the Akaike Information Criterion (AIC) using Eq.~\ref{eq:AIC}. Since both models include the parameter $m_\nu$, this is not treated as an additional degree of freedom in the comparison. Our analysis yields $\Delta\mathrm{AIC}=+0.7$, indicating weak evidence against the 4pDE model relative to the $\Lambda$CDM model for the combined dataset of Planck, DESI DR2 BAO, and Pantheon+.

\section{Conclusions}\label{sec:conclusion}

In this work, we revisited cosmological constraints on the total neutrino mass in the context of an extended four-parameter dark energy model, 4pDE, motivated by the possibility that late-time deviations from a cosmological constant may bias or relax neutrino-mass bounds derived under more restrictive assumptions. Using a combination of Planck 2018 temperature, polarization, and lensing measurements, DESI DR2 BAO distances, and the Pantheon+ supernova sample, we performed a comprehensive MCMC analysis with a modified version of \texttt{CLASS} interfaced through the \texttt{COBAYA} framework.

We find that the inclusion of DESI DR2 BAO significantly strengthens the constraints on the dynamical dark-energy equation-of-state parameters relative to earlier analyses of the same 4pDE model \cite{Sharma:2022ifr}. In particular, the high-precision distance measurements from DESI tighten the allowed ranges of the transition parameters $(a_{\rm t}, \Delta_{\rm de})$ and constrain the present-day value of the equation of state, $w_0$, in a manner consistent with the findings from DESI collaboration within the $w_0w_a$CDM framework \cite{DESI:2025zgx}. Moreover, the reconstructed equation of state shown in Figure~\ref{fig:reconstruction} does not exhibit a clear phantom crossing at $z \sim 0.5$, which supports the similar conclusion obtained from the DESI collaboration regarding a non-detection of a statistically significant phantom crossing \cite{DESI:2025zgx, DESI:2025fii}. This suggests that current data does not exclude an evolving equation of state that departs from a simple monotonic thawing or freezing evolution. 

Our analysis yields a relaxed $95\%$ C.L. upper limit on the total neutrino mass, $\sum m_\nu < 0.101,\text{eV}$, in contrast to the tighter $\Lambda$CDM bound of $\sum m_\nu < 0.0654,\text{eV}$. This relaxation is a direct consequence of the well-known degeneracy between $\sum m_\nu$ and the dark-energy equation of state: both massive neutrinos and evolving dark energy suppress the growth of structure and modify the expansion history, allowing variations in $w(z)$ to partially mimic neutrino-induced effects. At the same time, our 4pDE constraint is more stringent than the DESI DR2 $w_0w_a$CDM bound of $\sum m_\nu < 0.117$ eV, reflecting the increased constraining power of the combined dataset.

Although the 4pDE model provides a marginally better fit to the data ($\Delta\chi^2_{\rm min}=-7.3$), the Akaike Information Criterion yields $\Delta\mathrm{AIC}=+0.7$ from Equation \ref{eq:AIC}, indicating weak evidence against the extended model relative to $\Lambda$CDM. This highlights that while additional freedom in $w(z)$ affects degeneracies and parameter estimates, the current data do not statistically prefer the 4pDE framework.

Overall, our results underscore the importance of dark-energy modeling in deriving robust neutrino-mass constraints. As future DESI releases, Euclid, and Rubin LSST deliver sharper distance and growth measurements, the interplay between dynamical dark energy and neutrino physics will become increasingly important. The extended 4pDE model presented here offers a flexible framework for assessing these degeneracies and will be valuable for interpreting forthcoming precision datasets.

\section*{Data Availability}

The data sets used in this work are all publicly available. The modified codes used for this study may be made available upon reasonable request.

\section*{Acknowledgements}
We thank Ravi Kumar Sharma and William Giarè for their insightful discussions and assistance with running \texttt{COBAYA} for the last six months. AC and SD acknowledge the support by DST-DAAD Indo-German joint research collaboration grant DST/INT/DAAD/P-07/2023(G). AC and SD thank Heidelberg University for its kind hospitality and stimulating research environment, where part of the research work was carried out during the scientific visit. LA acknowledges support by DFG  under Germany's Excellence 
Strategy EXC 2181/1 - 390900948 (the Heidelberg STRUCTURES Excellence 
Cluster) and under Project  554679582 "GeoGrav: Cosmological Geometry
and Gravity with nonlinear physics".



\bibliographystyle{mnras}
\bibliography{ref} 








\bsp	
\label{lastpage}
\end{document}